\documentclass[aps,prl,twocolumn,superscriptaddress,groupedaddress]{revtex4}  
\usepackage{graphicx}  
\usepackage{dcolumn}   
\usepackage{bm}        
\usepackage{amssymb}   
\usepackage{amsmath}
\usepackage{subfigure}
\begin{document}

\title{Spatial Dispersal of Bacterial Colonies Induces a Dynamical Transition From Local to Global Quorum Sensing}
\author{Tahir I. Yusufaly}
\affiliation{Department of Physics and Astronomy, University of Southern California, Los Angeles, CA, 90089}
\author{James Q. Boedicker}
\affiliation{Department of Physics and Astronomy, University of Southern California, Los Angeles, CA, 90089}
\affiliation{Department of Biological Sciences, University of Southern California, Los Angeles, CA, 90089}
\date{\today}

\begin{abstract}
Bacteria communicate using external chemical signals called autoinducers (AI) in a process known as quorum sensing (QS). QS efficiency is reduced by both limitations of AI diffusion and potential interference from neighboring strains. There is thus a need for predictive theories of how spatial community structure shapes information processing in complex microbial ecosystems. As a step in this direction, we apply a reaction-diffusion model to study autoinducer signaling dynamics in a single-species community as a function of the spatial distribution of colonies in the system. We predict a dynamical transition between a local quorum sensing (LQS) regime, with the AI signaling dynamics primarily controlled by the local population densities of individual colonies, and a global quorum sensing (GQS) regime, with the dynamics being dependent on collective inter-colony diffusive interactions. The crossover between LQS to GQS is intimately connected to a tradeoff between the signaling network's latency, or speed of activation, and its throughput, or the total spatial range over which all the components of the system communicate. 

\end{abstract}
\maketitle

\begin{figure}
\includegraphics[scale=0.28]{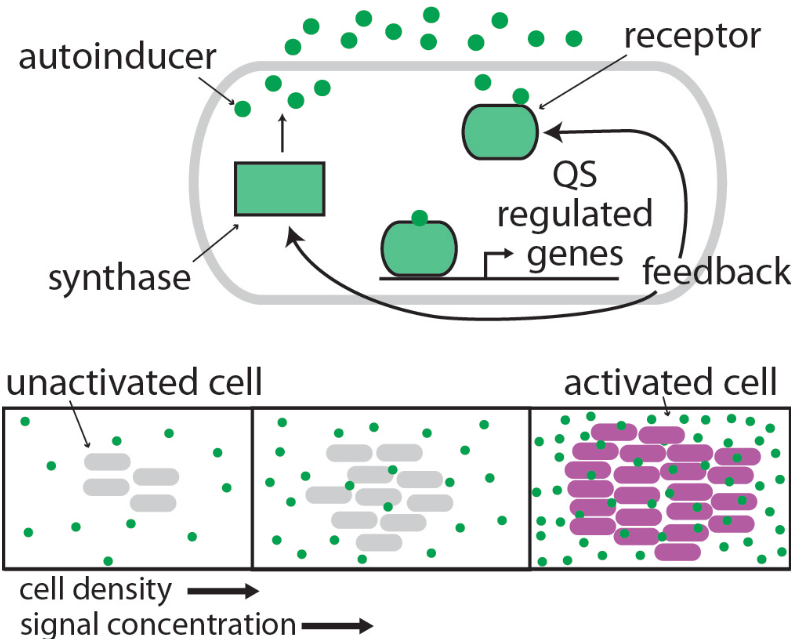}
\caption{The paradigmatic quorum sensing regulatory circuit consists of a synthase that produces autoinducer signals and a receptor that senses the local concentration of the same autoinducer. In the absence of a high density of bacteria, the synthase genes are expressed at a low basal level, secreting a small amount of autoinducer into the environment. Once the colonies have grown to beyond a critical population density, the collective concentration of autoinducer secreted is enough to activate quorum sensing.}
\label{Fig:AISignalingSchematic}
\end{figure}

Multicellular communities, such as colonies of bacteria, communicate with each other to coordinate changes in their collective group behavior. This communication usually takes the form of the production and secretion of extracellular signaling molecules called autoinducers (AI), as illustrated in Figure \ref{Fig:AISignalingSchematic}. Released autoinducers diffuse through the environment, and each cell senses the local concentration of signal to inform changes in gene regulation. This intercellular signaling network, known as quorum sensing (QS), is crucial for a wide array of important microbial processes, including biofilm formation, regulation of virulence and horizontal gene transfer \cite{QSReview, WilliamsReview, JayaramanReview}.

Decades of research have advanced our knowledge of QS, but several subtleties remain unresolved. In particular, AI signals may convey information about many aspects of the cellular network and local environment beyond simply the total number of cells in the system  \cite{WendellLimScience}. Far from being reducible to homogeneous, uniform density populations, microbial communities are typically characterized by high spatial heterogeneity \cite{Hibbing}. As a result, several new phenomena emerge due to crosstalk between spatially segregated populations \cite{Gore2016}. Consequently, it appears that AI molecules can be an indicator of increased local population density, and can also be proxies of other variables, such as population dispersal  \cite{Redfield2002, EffSens, PerezCrosstalk2011,BoedickerAngewChem,QSvsDS}.

In recent years, advances in the ability to experimentally probe the properties of cellular populations at the single-cell level \cite{QSExpt} have resulted in a growing community of theoretical physicists working to catalogue the different classes of collective behavior found in interacting communities of organisms \cite{LifeIsPhysics, NonEqPhysicsAndEvolution, BialekCriticality, JeffGorePaper, Boedicker2008}. This approach has already successfully yielded insight into a wide variety of ecological problems, with notable recent examples including the effects of invasion in cooperative populations \cite{EvolutionArrestsInvasion}, optimal foraging strategies in sheep herds \cite{SheepGrazing}, and the properties of microbial signal transduction networks \cite{LingChongYou,MehtaWingreen,LongWingreen} and their relationship to long-range spatial pattern formation \cite{DilanjiHagen2012,You2015}. However, to our knowledge, there have been few, if any, such studies on the specific question of exactly how spatiotemporal autoinducer profiles are influenced by the underlying heterogeneity of the microbial community that produces them.  

Accordingly, in this Letter, using a reaction-diffusion model, we identify how emergent spatiotemporal AI signaling patterns depend on community spatial structure. Via numerical simulations, we show that for a community of single-species bacteria, there is a transition in the activation dynamics as a function of community dispersal, or equivalently, the spatial heterogeneity of the cell density. At low dispersal, corresponding to a situation in which the cells aggregate into a few large colonies, the activation is triggered by the local size and density of the individual colonies, and therefore, can be described as local QS (LQS). At high dispersal, corresponding to a situation in which the cells are spread thinly among many small colonies, activation is instead a collective, global effect, mediated by mutual interactions between spatially disconnected colonies, and thus, this state is best described as one of global QS (GQS). 

\begin{figure}
\includegraphics[scale=0.5]{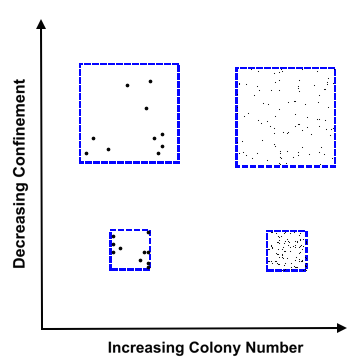}
\caption{In this study, we divide an initial population of $n_{0}$ cells into $N_{colonies}$ equally sized colonies, coexisting within a square region of length $L_{confinement}$. These two parameters, $N_{colonies}$ and $L_{confinement}$, serve as control parameters of our model. The number of colonies $N_{colonies}$ tunes the \textit{local} properties of individual colonies, while the confinement tunes the strength of \textit{global} inter-colony communication through molecular diffusion.}
\label{Fig:VaryingNumberofColonies}
\end{figure}

\begin{table}
\caption{\label{tab:Parameters} The representative set of parameters used in this work. Although these quantities possess substantial statistical uncertainty and condition-dependent variability, reported values in the literature typically fall within the same characteristic orders-of-magnitude.}
\begin{ruledtabular}
\begin{tabular}{lcr}
Quantity&Units&Value\\
\hline
Colony Cell Density ($\bar{c}_{colony}$) & cells/$\mu$m$^{3}$ & 1\footnote{Taken directly from \cite{EffSens}.}\\
Fold Change ($\textit{f}$) & Unitless & 6\footnote{Selected as a lower limit from the reported range of 5-fold to 300-fold change in \cite{TamsirBSIMLogic}, for numerical convenience.} \\
Signal Diffusivity ($\textit{D}$) & $\mu$m$^{2}$/sec & 160\footnote{Taken, with slight alteration for computational convenience, from \cite{TamsirBSIMLogic}.}\\
Signal Decay Rate ($\gamma$) & molecules/hr & 0.01\footnote{Taken directly from \cite{ChandlerIMSEJ}.} \\
Activation Threshold ($c_{AI_{crit}}$) & molecules/$\mu$m$^{3}$ & 100\footnote{Taken as a computationally convenient number corresponding to 160 nM concentration, an intermediate between the 10-70 nM reported in \cite{EmereniniPLOS} and the 450 nM reported in \cite{HagenTravelingWaves}.} \\
Basal Production Rate ($r_{b}$) & molecules/hr/cell & 500\footnote{Adapted from information in \cite{HagenTravelingWaves}. In particular, point production intensity is (0.43 nM/hr) $\times$ \textit{l}, where $l$ is the luminescence, ranging between 0 to 3000 (normalized), thus corresponding to an intensity range from about 0 to 1300 nM/hr. For a 1 $\mu$m$^{3}$ cell, this corresponds to approximately 0 to 1000 molecules per hour per cell, with the midpoint value of 500 thus being a reasonable estimate of a typical rate. } \\
Doubling Time & mins & 30\footnote{Rounded up from mean values reported in \cite{Powell1956}.} \\
Initial No. of Cells ($n_{0}$) & 10$^{3}$ cells & 7 \\
Max. No. of Cells ($n_{max}$) & 10$^{3}$ cells & 7000 \\
\end{tabular}
\end{ruledtabular}
\end{table}

The dynamics of the autoinducer concentration $c_{AI}$ are governed by the spatiotemporal cell density profile $c_{cell}$ via the reaction-diffusion equation
\begin{equation}
(\frac{\partial}{\partial t} - D \nabla^{2} + \gamma) c_{AI} = r(c_{AI}) c_{cell}(\vec{x}, t) \ ,
\label{eq:RxnDiffEquation}
\end{equation}
where $D$ and $\gamma$ are respectively the AI signal diffusivity and decay, and $r$ is the local, AI-concentration-dependent signal production rate of the cells. In this work, we restrict ourselves to a regime where the timescale of sigmoidal quorum sensing activation is taken to be much faster than the timescale for signal production and diffusion, which implies that in the limit of coarse-grained, long-time integrations, the activation can be approximated by an instantaneous all-or-none step function (for a specific recent study analyzing the range of validity of this approximation in quorum sensing systems, we refer the reader to \cite{StepActivation}, and for further validation data comparing results of the Heaviside activation to a more realistic Hill curve, we refer to Figure 5(a) in the Appendix). Thus, we can approximate the activation to be instantaneous, modeling it via the Heaviside theta function $\theta$:
\begin{equation}
r(c_{AI}) = r_{b}(1 + f \theta(c_{AI}(\vec{x}, t) - c_{AI_{crit}})) \ .
\label{eq:RateActivationEquation}
\end{equation}

Here, $r_{b}$ is the basal AI production rate of the cells in the absence of any activation, while $c_{AI_{crit}}$ is the threshold AI concentration for cells to transition to an activated state, with a fold change $f$ ratio increase in synthase gene expression, and thus AI production rate. 

Additionally, we take the cell density $c_{cell}$ to be a Boolean function, adopting either a constant average $\bar{c}_{colony}$ for points occupied by a colony, or 0 otherwise. With this setup, the spatiotemporal density of the cellular community $c_{cell}$ can be interpreted as an input driving signal that generates the spatiotemporal AI profile $c_{AI}$ as an output response signal. 

We design and perform a series of computational experiments to isolate the effects of spatial dispersal on AI signaling. At time $t = 0$, we instantaneously colonize a quasi-2D region of space with an initial population of $n_{0}$ cells, and have the cells exponentially grow until they have multiplied to a final total carrying capacity value $n_{max}$, at which point the growth is saturated. 

We perform several replicates of this simulation. In the first replicate, the cells are perfectly localized into a circular colony at the center of the quasi-2D region, corresponding to a situation of minimal dispersal, with only one homogeneous colony and no spatially distributed subpopulations. The cells produce, sense, and activate AI signals undergoing diffusion and decay according to equations (\ref{eq:RxnDiffEquation})-(\ref{eq:RateActivationEquation}), with the representative parameter values shown in Table \ref{tab:Parameters}. The system is numerically integrated to yield a discretized output spatiotemporal AI signaling profile $c_{AI}(\vec{x}, t)$. Calculations are performed with the aid of the BSIM software package \cite{BSIM}; for additional details on the setup of the simulation, we refer the reader to the Appendix.

\begin{figure}
\subfigure[]{\includegraphics[scale=0.12]{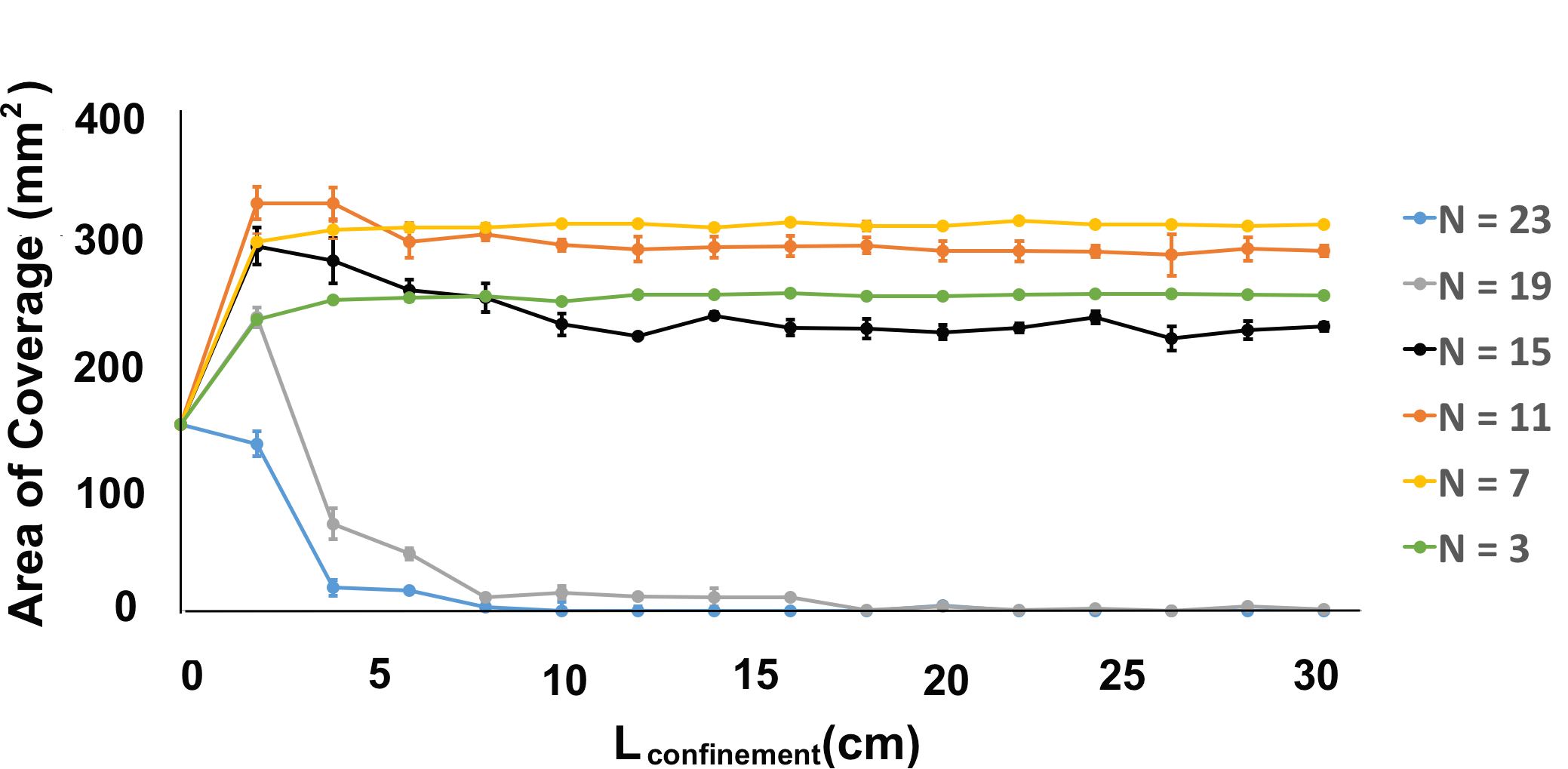}
\label{subfig:TenHourActivationLevels}}
\subfigure[]{\includegraphics[scale=0.12]{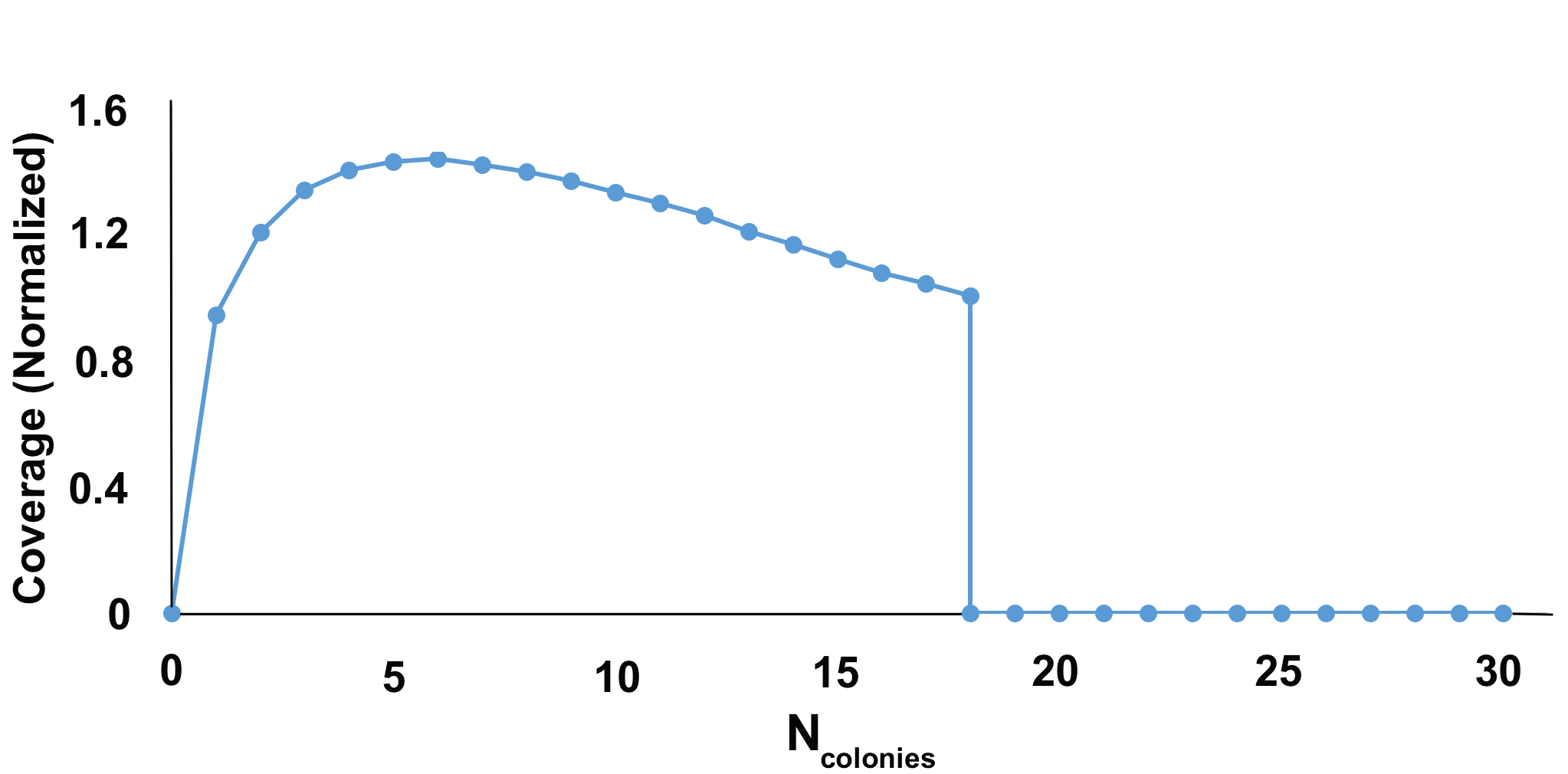}
\label{subfig:AsymptoticCoverageUnconfinedLimit}}
\caption{(a) The average coverage, or area of space with above-threshold signal concentration, at a representative time point of ten hours, plotted as a function of the dimension of confinement $L_{confinement}$ for varying numbers of colonies $N_{colonies}$. Residuals represent standard errors over five simulation replicates each. (b) Asymptotic coverage level as a function of $N_{colonies}$ in the limit of deconfined, or infinitely separated colonies. There is a clear transition from non-zero to zero levels between $N_{colonies} = 17$ and $18$, illustrating a crossover from local quorum sensing (LQS) to global quorum sensing (GQS).}
\label{Fig:SpatiotemporalSignalingDynamics}
\end{figure}

For subsequent replicates of the simulation, we keep these details the same, except that we divide the original single colony into various discrete numbers of colonies $N_{colonies}$, and confine the colonies to occupy a specified square area with side-length $L_{confinement}$. Then, different sized confinement squares tune the packing density of the colonies, or the ratio of the total area of all colonies to the area in which they are confined. This setup is displayed in Figure \ref{Fig:VaryingNumberofColonies}. In all cases, the initial number of cells $n_{0}$ and final number of cells at full growth $n_{max}$ are the same as in the single-colony simulation. Consequently, increasing values of $N_{colonies}$ gradually tune the system from a state of a few large colonies, to one of many smaller colonies, with a corresponding increase in the concentration of interacting patches of spatially extended subpopulations. Additionally, increasing areas of confinement gauges the importance of inter-colony interactions, because as the packing density decreases, so does the average inter-colony spacing, and thus, the magnitude of diffusive crosstalk between spatially separated colonies. 

\begin{figure}
\subfigure[]{\includegraphics[scale=0.15]{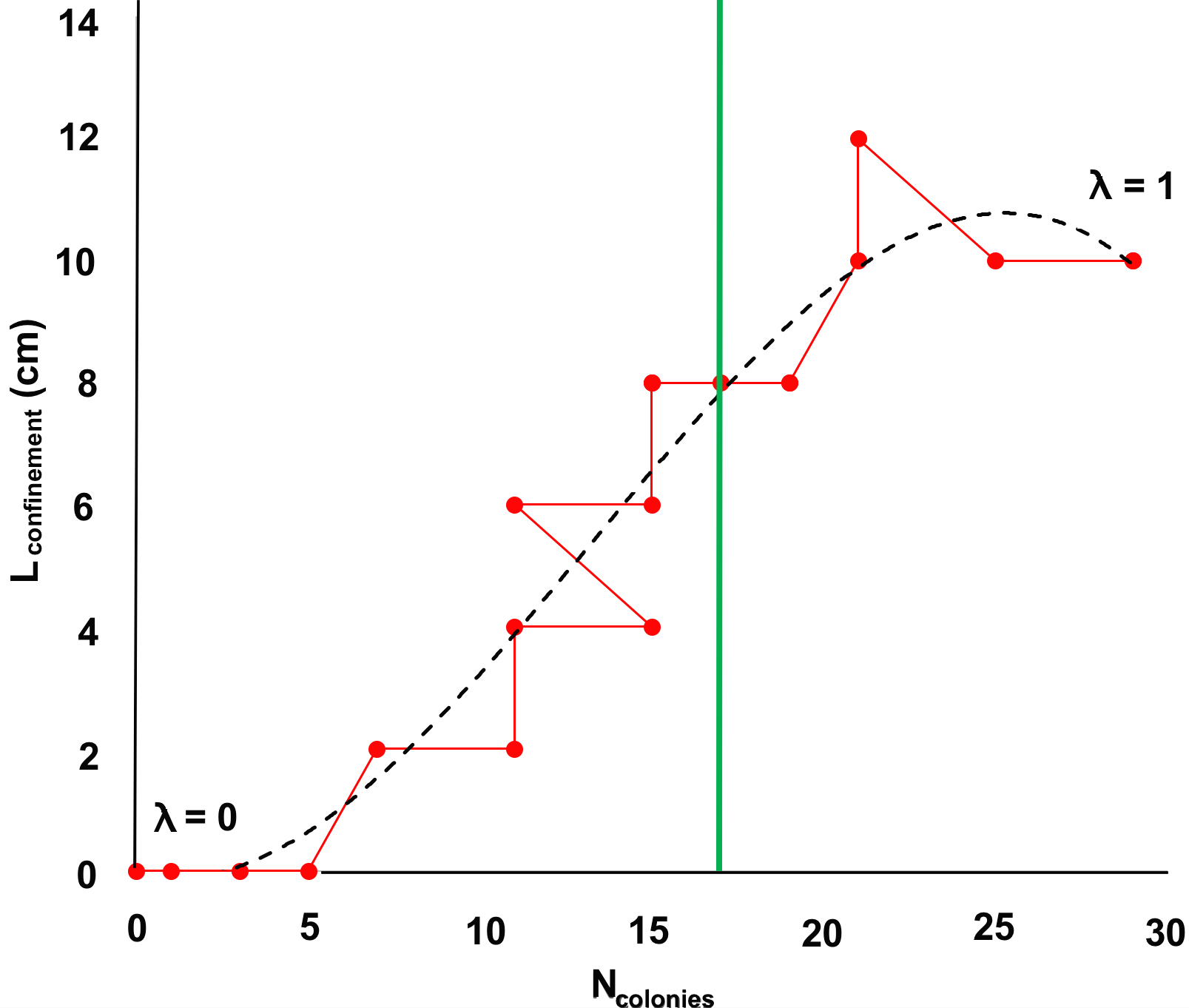}}
\subfigure[]{\includegraphics[scale=0.09]{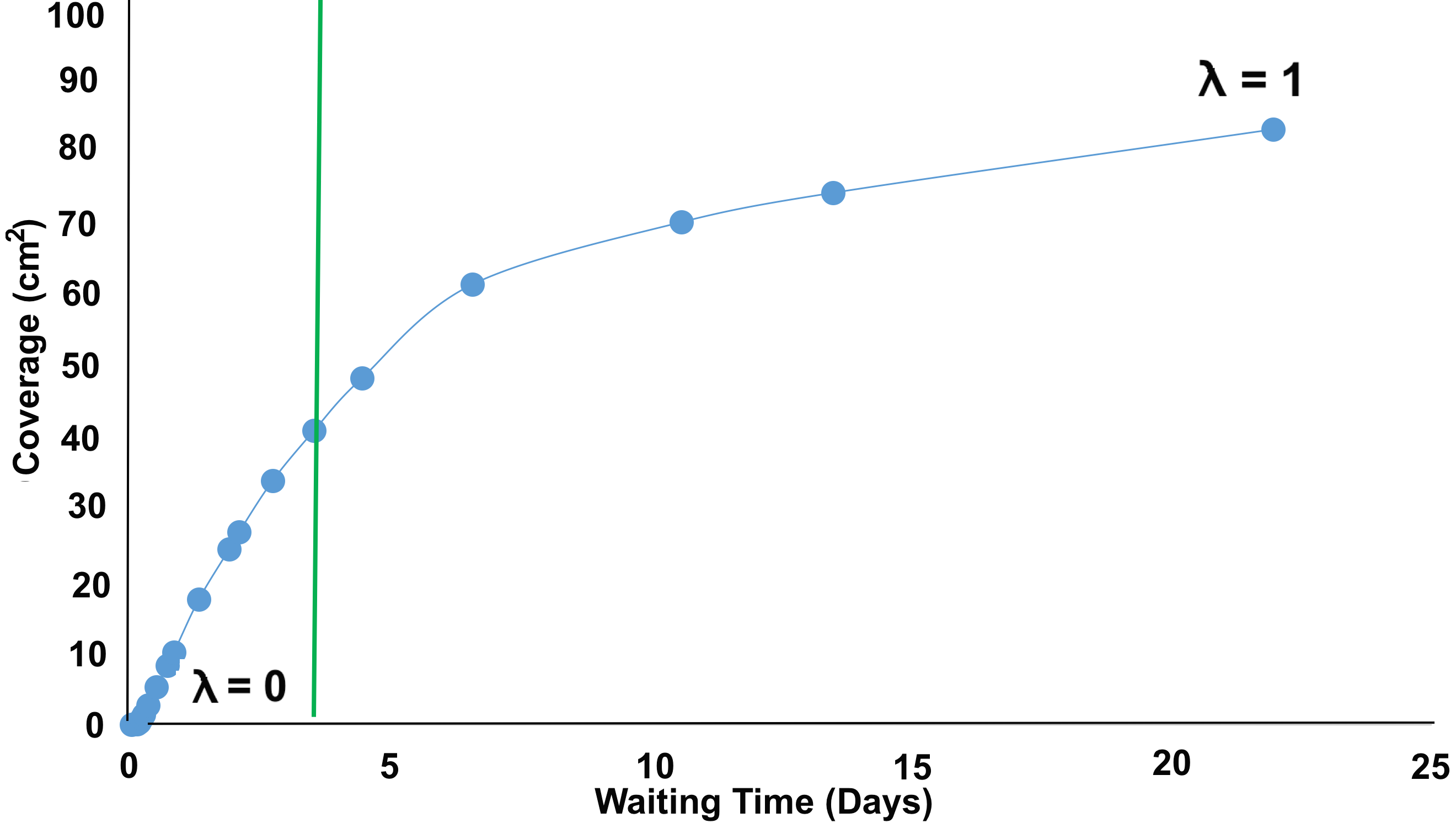}}
\caption{(a) A Pareto optimal curve of control parameters ($N_{optimal}(\lambda)$, $L_{optimal}(\lambda)$), parameterized by an affine parameter $\lambda$ ranging from 0 to 1. Each value of $\lambda$ maps on to an associated Pareto point in ($N_{colonies}$, $L_{confinement}$) parameter space, such that, for a given waiting time from the moment of initial colonization $\tau(\lambda)$ the overall coverage $C$ (defined in Figure \ref{Fig:SpatiotemporalSignalingDynamics}) at $\tau$ is maximized (within the resolution of the numerical simulations) - quantitatively, $C_{max}$($\lambda$) = $C$($\tau$($\lambda$), $N_{optimal}$($\lambda$), $L_{optimal}$($\lambda$)) $\geq$ $C(\tau(\lambda), N_{colonies}, L_{confinement})$ for all $N_{colonies}$ and $L_{confinement}$.  The red dots correspond to the observed discrete values of ($N_{colonies}$, $L_{confinement}$), while the black dotted line is a resulting best-fit interpolation curve. Displayed below this optimal curve ($N_{optimal}(\lambda)$, $L_{optimal}(\lambda)$) is (b) a plot of the corresponding maximal coverage $C_{max}$($\lambda$) vs. the waiting time $\tau$($\lambda$). The green lines indicate a crossing from an LQS to GQS regime of quorum sensing, resulting in a rapid onset of `diminishing returns' in throughput, relative to the latency cost.}
\label{Fig:SelfActivationCrossOver}
\end{figure}

We define a region of space as being activated if its concentration of AI signal is above the quorum sensing threshold $c_{AI_{crit}}$. Figure \ref{subfig:TenHourActivationLevels} illustrates the fraction of space that gets activated, at a sample intermediate time of ten hours, for the different spatial geometries studied. At low colony number $N_{colonies}$, corresponding to a small number of highly clustered colonies, the system is in a regime of local quorum sensing (LQS), where regions near the individual colonies receive enough locally produced signal to enable rapid activation, in a manner that is robust to the activity of neighboring colonies. This robustness can be seen by observing how the activation of the system changing as $L_{confinement}$ increases, corresponding to further separation between colonies. We see that for $N_{colonies} < N_{c} = 18$, the activation does not vanish but approaches a finite limiting value. Meanwhile, for $N_{colonies} > N_{c}$, the system is in a regime of global quorum sensing (GQS), where activation is necessarily communal in nature, triggered by the collective sharing of AI signal between all colonies. Here, as $L_{confinement}$ increases, the activation level vanishes because individual colonies are too small to self-activate without crosstalk from neighboring colonies. These arguments are further corroborated in Figure \ref{subfig:AsymptoticCoverageUnconfinedLimit}, which shows asymptotic long-time coverage in the deconfined limit, $L_{confinement} \rightarrow \infty$, where the colonies are completely non-interacting. In this regime, it is even more apparent that there is a discontinuous transition in activation at $N_{c}$. In the Appendix, our numerical results are compared to an analytical mean-field model which captures the essential physics governing the transition.

Furthermore, working in non-dimensionalized units (see Appendix for more details; for the purposes of continuity, we stick to conventional units throughout the main manuscript) demonstrates that, despite the significant number of input quantities in the model, the qualitative behavior of the system is invariant to most of the detailed parameter values, and instead, only depends on the basal activity level of colonies, as determined by $r_{b}$, the fold change $f$ upon activation, and the maximal cell number $n_{max}$. The existence of a critical threshold $N_{c}$ separating these two regimes is independent of the fold-change $f$, as activation of an isolated colony depends only on its own basal activity and geometry, and is not influenced by any possible activated neighboring colonies. In addition, variations in the total cell number $n_{max}$ and basal activity level $r_{b}$ simply shift $N_{c}$. In particular, increasing the value of $n_{max}$ simply increases $N_{c}$ by increasing the number of divisions required to fine-grain individual colonies to a size below the self-activation limit. Meanwhile decreasing $r_{b}$ simply decreases $N_{c}$ by decreasing the basal intensity per unit area, and thus the total basal intensity for a given colony radius.

This transition between LQS to GQS is related to tradeoffs between the signaling network's latency, or speed of activation, and its throughput, or the total spatial range over which all the components of the system communicate \cite{LatencyVsThroughput}.  Communities in the LQS regime have a reduced time to activation, but are restricted to short-range communication.  The large colony size leads to production of signal near the colony center beyond what is needed for local activation.  The slow rate of signal diffusion coupled with larger intercolony distances does not allow the signal to effectively communicate with neighboring colonies.  On the contrary, although communities in the GQS regime take longer to initiate activity, the broadly distributed colonies of smaller size enable long-range communication throughout the network of cells.

These tradeoffs are illustrated in Figure \ref{Fig:SelfActivationCrossOver} describing a Pareto optimal contour in parameter space, with each point on the contour corresponding to a choice of parameters that, for a given waiting time from the moment of initial colonization $\tau$, optimizes the overall coverage at that time, which we denote $C(\tau)$. The results show that if one is primarily interested in rapid initiation, it is preferable to be in the LQS regime, but that if one is interested in having the maximum possible fraction of space receive the signal in a reasonable amount of time, it is preferable to be near the GQS regime. 

These characteristics of the optimal contour are, to a large degree, independent of slight variations in parameter values. As has already been discussed, changes in $n_{cells}$ and $r_{b}$ simply shift $N_{c}$, and thus, only show up as smooth deformations of the optimal contour, with the qualitative form of the peak, as well as its dynamic variation, still intact. Furthermore, slight variations in the remaining free parameter, the fold change $f$ simply amount to changes in the steepness of the contour, by stretching or compressing the magnitude of separation between active and inactive coverage levels. However, the fact remains that in the GQS regime, $N_{colonies} > N_{c}$, there comes a point of `diminishing returns', where coverage eventually decreases as $L_{confinement}$ increases, due to activation depending on the distance between neighboring colonies. For additional numerical evidence validating these physical arguments, we refer the reader to Figures 5(b) and 5(c) of the Appendix.

In summary, the results of this work clearly point to the prediction of an experimentally observable transition between LQS and GQS behavior. This transition is observed to be intimately connected to a tradeoff between short-term latency, optimized in the LQS regime, and long-term throughput, optimized in the GQS regime. The qualitative results are independent of specific parameter values, being primarily a consequence of how coverage dynamics scale as colony number and confinement are varied. A natural follow-up to this work would be exploring how the behavior of the tradeoffs is influenced by heterogeneity, for example, due to the presence of multiple species and single-cell stochasticity.

\textit{Acknowledgments}: This work was funded by the Office of Naval Research under Grant N00014-15-1-2573. We thank Sylvian Hemus for early help with this project, and Osman Kahraman and Christoph Haselwandter for reading and commenting on the manuscript. We would also like to thank all the referees who reviewed our manuscript, whose suggestions greatly improved the work.

\section{Appendix}

\subsection{Continuous Formulation of Reaction-Diffusion Equations}
The dynamics of our system are governed by the reaction-diffusion equation
\begin{equation}
(\frac{\partial}{\partial t} - D \nabla^{2} + \gamma) c_{AI} = r(c_{AI}) c_{cell}(\vec{x}, t) \ ,
\label{eq:RxnDiffEquation}
\end{equation}
where $D$ and $\gamma$ are respectively the AI signal diffusivity and decay, and $r$ is the local, AI-concentration-dependent signal production rate of the cells, given by
\begin{equation}
r(c_{AI}) = r_{b} (1 + f \theta(c_{AI}(\vec{x}, t) - c_{AI_{crit}}) ) \ .
\label{eq:RateActivationEquation}
\end{equation}
Here, $r_{b}$ is the basal AI production rate of the cells in the absence of any activation, while $c_{AI_{crit}}$is the threshold AI concentration for cells to transition to an activated state, with a fold change $f$ ratio increase in synthase gene expression, and thus AI production rate.. We take the activation to be instantaneous, modeling it via the Heaviside theta function $\theta$, which is a reasonable approximation to the smoother Hill function as long as the timescale of the activation reaction is much faster than that of diffusion or signal production. With this setup, the time-dependent cell density profile $ c_{cell}(\vec{x}, t) $ serves as a tunable input driving signal, generating an output $c_{AI}(\vec{x}, t)$. 

As discussed in the text, the cell density is divided into a discrete number of equally sized building-block pieces centered at an equivalent number of colony center points, with each building-block representing a separate colony. The building-blocks are taken to have a given cell density $\bar{c}_{colony}$, representing the number of cells per unit area in a colony, and a time-dependent radius $R(t)$, representing colony growth. Thus, a given building-block density profile, for a colony centered at $\vec{x}_{0}$, is
\begin{equation}
c_{colony}(\vec{x}, t; \vec{x}_{0}, R(t)) = 
\begin{cases} 
      \bar{c}_{colony} & |\vec{x} - \vec{x}_{0}| \le R(t) \\
      0 & |\vec{x} - \vec{x}_{0}| > R(t) . \\
\end{cases} 
\label{eq:ColonyProfile}
\end{equation}
With this notation, if we confine our cells to a square area $A_{confinement}$ = $L_{confinement} \times L_{confinement}$, then a general cell density profile $c_{cell}$ can be written as a linear superposition of all colonies in the system
\begin{equation}
c_{cell}(\vec{x}, t) = \sum_{i = 1}^{N_{colonies}} c_{colony}(\vec{x}, t; R(t), \vec{x}_{0i}(L_{conf.}) )
\label{eq:BuildingBlockDecomposition}
\end{equation}
where $N_{colonies}$ is the total number of colonies, $\{ \vec{x}_{0i} \}$ is the set of colony centers and $R(t)$ is the colony radii, which is constant for all colonies. As shown in equation (\ref{eq:BuildingBlockDecomposition}), the possible center positions are functionally constrained by the confinement area. We will discuss the choice of the $\{ \vec{x}_{0i} \}$ momentarily, but first we address determination of $R(t)$. We proceed by noting that the total number of cells $n_{cells}$ in the system starts at a given initial number $n_{0}$, then exponentially grows to a given maximum level $n_{max}$, with a given time constant $t_{1/2}$, at which point the cells immediately stop dividing. The time $t_{max}$ at which the population reaches this maximum level can easily be calculated by setting $n_{max} = n_{0} e^{t_{max} / t_{1/2}}$, which gives 
\begin{equation}
t_{max} = t_{1/2} \ \text{ln}(n_{max} / n_{0}) \ .
\label{eq:tmax}
\end{equation} 

Thus, the explicit time-dependent profile for the total number of cells is 
\begin{equation}
n_{cells}(t) = 
\begin{cases} 
      n_{0} e^{t / t_{1/2}} & 0\leq t \leq t_{max} \\ 
      n_{max} & t_{max} < t \ .
   \end{cases}
\label{eq:TotalCells}
\end{equation}

To relate this to colony radii, we note that the total cell density profile $n_{cell}$, for all replicates, must satisfy the normalization condition
\begin{equation}
\int_{-L/2}^{L/2} \int_{-L/2}^{L/2} d^{2} \vec{x} \ c_{cell}(\vec{x}, t) = N_{cells} (t) \ .
\label{eq:Normalization}
\end{equation}

Thus, if the total number of cells is equally divided between $N_{colonies}$ number of colonies, each with uniform cell density $\bar{c}_{colonies}$, then it follows that the radius $R(t)$ of each colony must be equal to
\begin{equation}
R(t) =\frac{1}{N_{colonies}} \sqrt{\frac{n_{cells}(t)} {\pi \bar{c}_{colonies}}} .
\label{eq:Roft}
\end{equation}
We note that it follows immediately from this result that the radius of a single colony at full growth is given by
\begin{equation}
R_{colony} = \frac{1}{N_{colonies}} \sqrt{\frac{n_{max}} {\pi \bar{c}_{colonies}}} .
\label{eq:Rcols}
\end{equation}

With this information, we now proceed to identify a range of possible colony centers - in particular, we would like to calculate, for a given number of colonies $N_{colonies}$, each with radius at full growth $R_{colony}$, and confined to an $L_{confinement} \times L_{confinement}$ square, the probability distribution over the $N_{colonies}$-dimensional random vector of colony centers $\vec{X}_{0} = (\vec{x}_{01}, \vec{x}_{02}, ..., \vec{x}_{0N_{colonies}}$). The range of allowed center positions is set by two constraints: 1) no center can be closer than a distance $R_{colony}$ to the edge of the square of confinement, to ensure that no cells leave the confinement boundary, and 2) colony centers must be separated by a distance of at least $2R_{colony}$ in order to avoid overlap. Thus, the probability distribution $p(\vec{X}_{0})$ is simply a uniform distribution over all colony center combinations that remain within the prescribed boundaries and which do not contain pairs of points closer than the cutoff separation. While an analytical form of this distribution is, in general, untenable, it is straightforward to implement numerically.

\subsection{Non-Dimensionalization}

We can reduce the total number of independent parameters by transforming to unitless dimensions of space, time, molecular and cellular concentrations. We start with our governing equation,
\begin{equation}
(\frac{\partial}{\partial t} - D \nabla^{2} + \gamma) c_{AI} = r(c_{AI}) c_{cell}(\vec{x}, t) \ ,
\label{eq:RxnDiffEquation}
\end{equation}
with
\begin{equation}
r(c_{AI}) = r_{b} (1 + f \theta(c_{AI}(\vec{x}, t) - c_{AI_{crit}}) ) . 
\label{eq:RateActivationEquationWithFFunction}
\end{equation}
If we make a change of variables to unitless quantities, via the prescription
\begin{equation}
(\vec{\tilde{x}}, \tilde{t}, \tilde{c}_{AI}, \tilde{c}_{cell}) = (\sqrt{\frac{\gamma}{D}} \vec{x}, \gamma t, \frac{c_{AI}}{c_{AI_{crit}}}, \frac{c_{cell}}{\bar{c}_{colony}}) \ ,
\end{equation}
then equation (\ref{eq:RxnDiffEquation}) adopts the non-dimensionalized form
\begin{equation}
(\frac{\partial}{\partial \tilde{t}} - \tilde{\nabla}^{2} + 1) \tilde{c}_{AI} = \tilde{r}(\tilde{c}_{AI}) \tilde{c}_{cell}(\vec{\tilde{x}}, \tilde{t}) \ .
\end{equation}
Thus, $\tilde{c}_{cell}$ is now a unitless source field that adopts discrete Boolean values of either 0 (corresponding to no cells present at a particular point in space and moment in time) or 1 (corresponding to cells being present with density $\bar{c}_{colony}$). Furthermore, the activation function (\ref{eq:RateActivationEquationWithFFunction}) has been transformed into the rescaled, unitless form
\begin{equation}
\tilde{r}(\tilde{c}_{AI}) = (\frac{r_{b} \bar{c}_{colony}}{\gamma c_{AI_{crit}}}) (1 + f \theta(\tilde{c}_{AI} - 1)) .
\end{equation}
The first set of parameters in parentheses on the right-hand side is nothing more than the unitless steady-state basal intensity of a point source of density $\bar{c}_{colony}$, as can be seen by setting the rate of production equal to the rate of decay, $r_{b} \bar{c}_{colony} = \gamma c_{SS}$, from which it is easily shown that
\begin{equation}
\tilde{c}_{SS} = \frac{c_{SS}}{c_{AI_{crit}}} = \frac{r_{b} \bar{c}_{colony}}{\gamma c_{AI_{crit}}} .
\end{equation}
Summarizing, we have shown that, through judicious choice of units, the behavior of the system depends on significantly fewer parameters than we started out with - in particular, only the basal activity level $\tilde{c}_{SS}$ and the fold change upon activation $f$ tune the output autoinducer response $\tilde{c}_{AI}(\vec{\tilde{x}}, \tilde{t})$ for a given input cell source profile $\tilde{c}_{cell}(\vec{\tilde{x}}, \tilde{t})$.

\subsection{Discretization and Computation}
Discretization of the continuous reaction-diffusion system for numerical computation is fairly straightforward. We perform calculations in a simulation box of dimensions $L^{box}_{x} \times L^{box}_{y} \times L^{box}_{z}$ where $L^{box}_{x}$ and $L^{box}_{y}$ are set to be $\Delta L$ = 5 cm greater than the confinement length $L^{box}_{x} = L^{box}_{y} = L_{confinement} + \Delta L$, in order to reduce finite-size numerical artifacts, and $L^{box}_{z} = 1 \ \mu m$. Continuous functions are discretized into an $N^{grid}_{x} \times N^{grid}_{y} \times N^{grid}_{z}$ mesh of rectangular units, each unit consequently having dimensions $\frac{L^{box}_{x}}{N^{grid}_{x}} \times \frac{L^{box}_{y}}{N^{grid}_{y}} \times \frac{L^{box}_{z}}{N^{grid}_{z}}$. Mesh units are chosen such that $\frac{L^{box}_{x}}{N^{grid}_{x}} = \frac{L^{box}_{y}}{N^{grid}_{y}} = 500 \mu m$, with $N^{grid}_{z} = 1$. In addition, time is discretized into windows of length $\delta t = 60 \ \text{minutes}$, which for a simulation over a time period $t^{sim} = 30 \ \text{days}$ results in $t^{sim}/ \delta t$ evenly sampled time points. Thus, the continuous functions $c_{AI}(\vec{x}, t)$ and $c_{cell}(\vec{x}, t)$ are each replaced by an $N^{grid}_{x} \times N^{grid}_{y} \times N^{grid}_{z} \times \delta t$ array of numbers in the $x$, $y$, $z$ and $t$ coordinates, respectively. In the quasi-2d limit studied in this work, where $N^{grid}_{z} = 1$, and thus we can refer to the discretized version of a function by $f(\vec{x}, t) \rightarrow f[i_{x}, i_{y}, i_{t}]$, where $i_{x}$ and $i_{y}$ represent the mesh of points in the $x$ and $y$ directions, $i_{t}$ represents a time point, and it is implicitly understood that we are fixed at the sole lattice point in the $z$ direction.

The input cell density profile array is constructed as a superposition of the corresponding colony building-block arrays, as shown in Equation (\ref{eq:BuildingBlockDecomposition}), with the colony centers chosen at random to avoid overlap and remain within confinement boundaries, as described previously. Five different replicates were run for each value of $N_{colonies}$ and $L_{confinement}$, and the mean and standard error are calculated (note that in the limit of either a single colony or unconfined, infinitely separated colonies, there is no variation between different replicates, so the standard error is trivially zero). For explanatory purposes, it suffices to describe how to construct a single colony. 

Each colony is discretely build up the density using units such that at time $t = 0$, one point has been added, at the mesh unit associated with the corresponding colony center. Then, the colony will grow over the time window $0 < t < t_{max}$, via the sequential addition of discrete points, until $N_{max}/N_{0}$ points have been added. 

When building up a system consisting of $N_{colonies}$ total colonies, in order to ensure a final total intensity of cell number $\frac{n_{cells}(t)}{N_{colonies}}$ in each individual colony, we build up any individual points in discrete intensity values of $\frac{n_{0}}{n_{max}}\frac{n_{cells}(t)}{N_{colonies}}$. 

With this information, then, for a colony, centered at a point $(x_{0}, y_{0})$, the coordinates of each successive point $(x, y)$ are parametrized by a discrete Archimedean spiral,
\begin{equation}
\begin{aligned}
(x[n], y[n]) &= (x_{0}, y_{0}) + \frac{1}{N_{colonies}} \times \sqrt{n} \\
& \times (\text{Cos}( 2 \pi \sqrt{n}), \text{Sin}(2 \pi \sqrt{n})) \ \mu \text{m}
\label{eq:ArchSpiral}
\end{aligned}
\end{equation}
where $n = 1, 2, ..., \frac{n_{max}}{n_{0}}$. This explicit form of the parametrization guarantees that upon completion of growth, the radius of the spiral disk is approximately equal to the desired maximum radius $R_{max}$. This completes the necessary information required to construct a density profile for individual colonies, and by superposition, entire cellular communities.

\subsection{Numerical Tests of Model and Parameter Robustness}
As discussed earlier, nondimensionalization reduces the number of independent parameters to three: the basal rate $r_{b}$, the total number of cells $n_{max}$ and the fold change $f$. Additionally, the major assumption of our model is approximation a smooth sigmoidal activation function function with a Boolean Heaviside-step function. Specifically, a more realistic representation of the activation is given by a Hill function with Hill coefficient $b$:
\begin{equation}
r_{Hill}(c_{AI}) = r_{b} (1 + \frac{f}{1 + (\frac{c_{AI}(\vec{x}, t)}{c_{AI_{crit}}})^{-b}}) \ .
\label{eq:RateActivationEquationHill}
\end{equation}
In the limit that $b \rightarrow \infty$, equation (\ref{eq:RateActivationEquationHill}) reduces to the numerically simpler approximation of equation (\ref{eq:RateActivationEquation}).

A change of cell number $n_{max}$ will simply shift the number of colony divisions required for individual colonies to reach the LQS/GQS threshold. In particular, if with $n_{max}$ cells, the critical colony number is $N_{crit}$, corresponding to a point where each colony has $\frac{n_{max}}{N_{crit}}$ cells individually. Then, if we scale $n_{max}$ by a factor $a$, individual colonies will now simply reach the critical cell number per colony at a scaled $a N_{crit}$.

\begin{figure}
\subfigure[]{\includegraphics[scale=0.175]{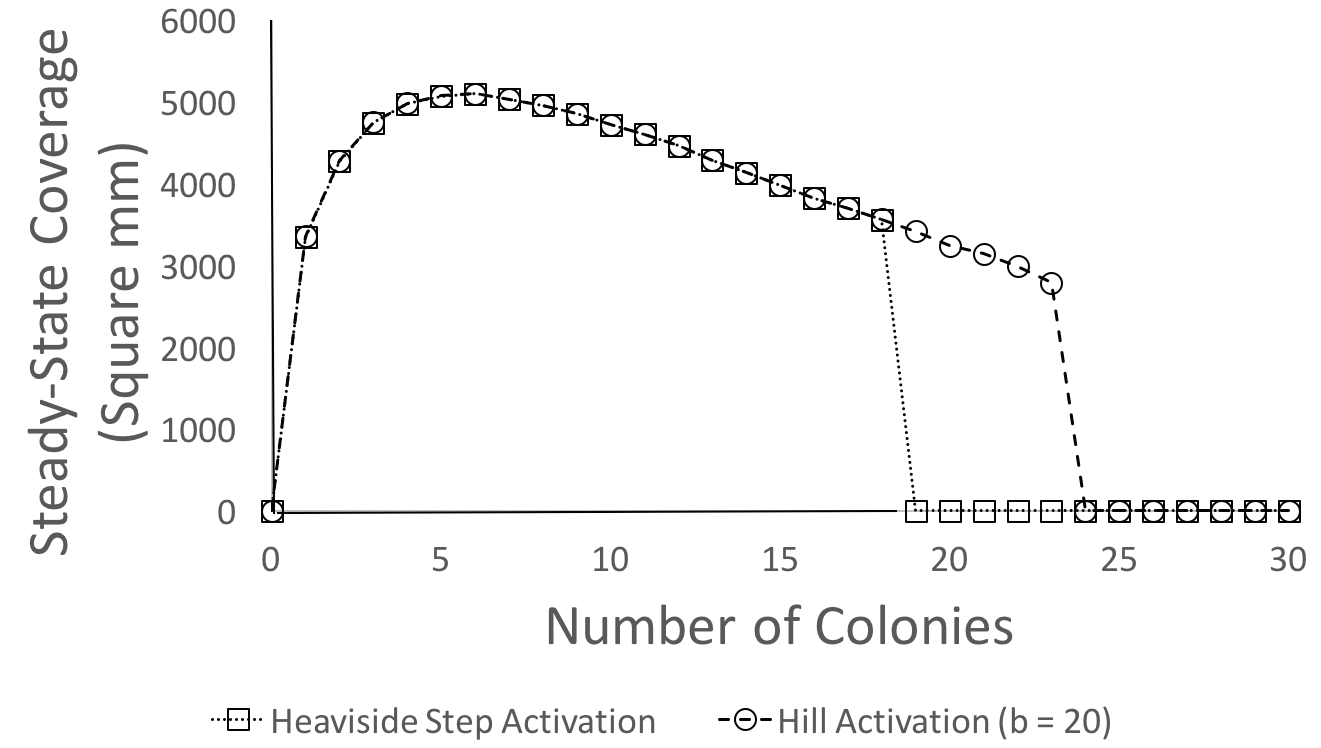}
\label{subfig:HillRobustness}}
\subfigure[]{\includegraphics[scale=0.175]{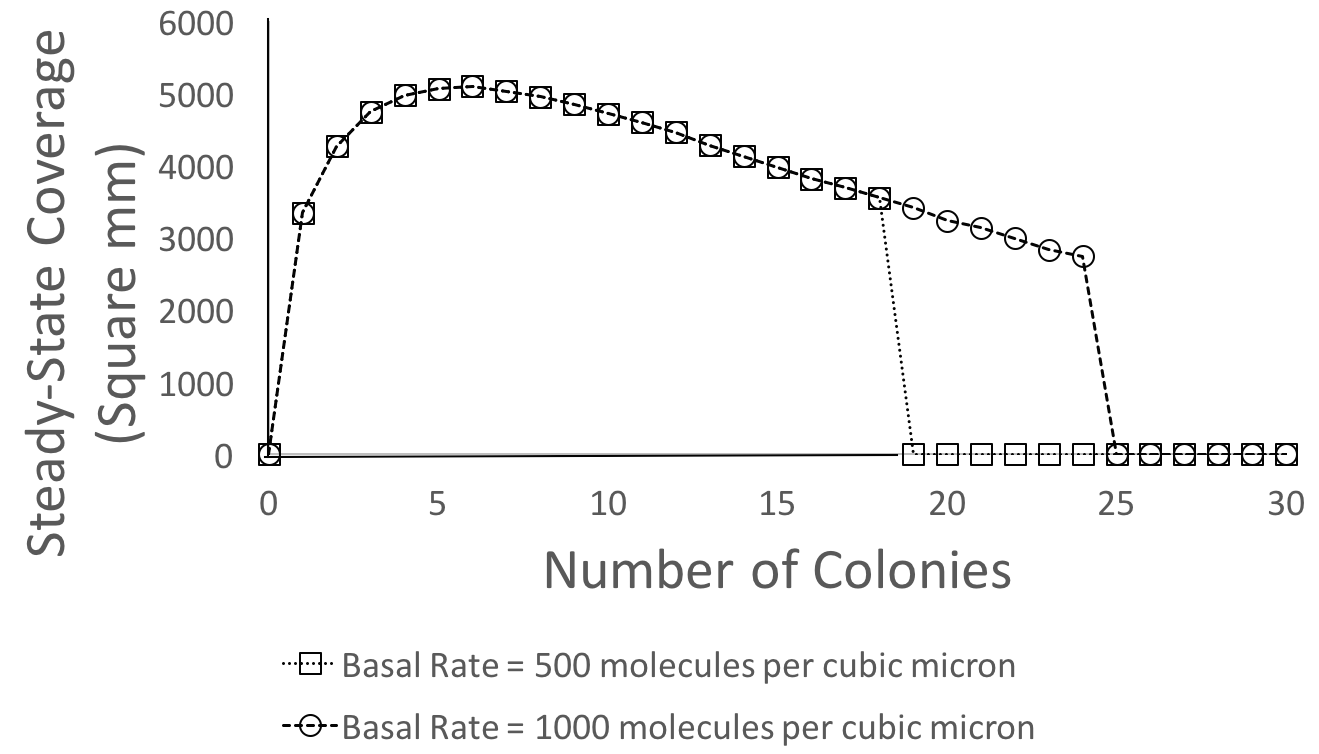}
\label{subfig:BasalRateRobustness}}
\subfigure[]{\includegraphics[scale=0.175]{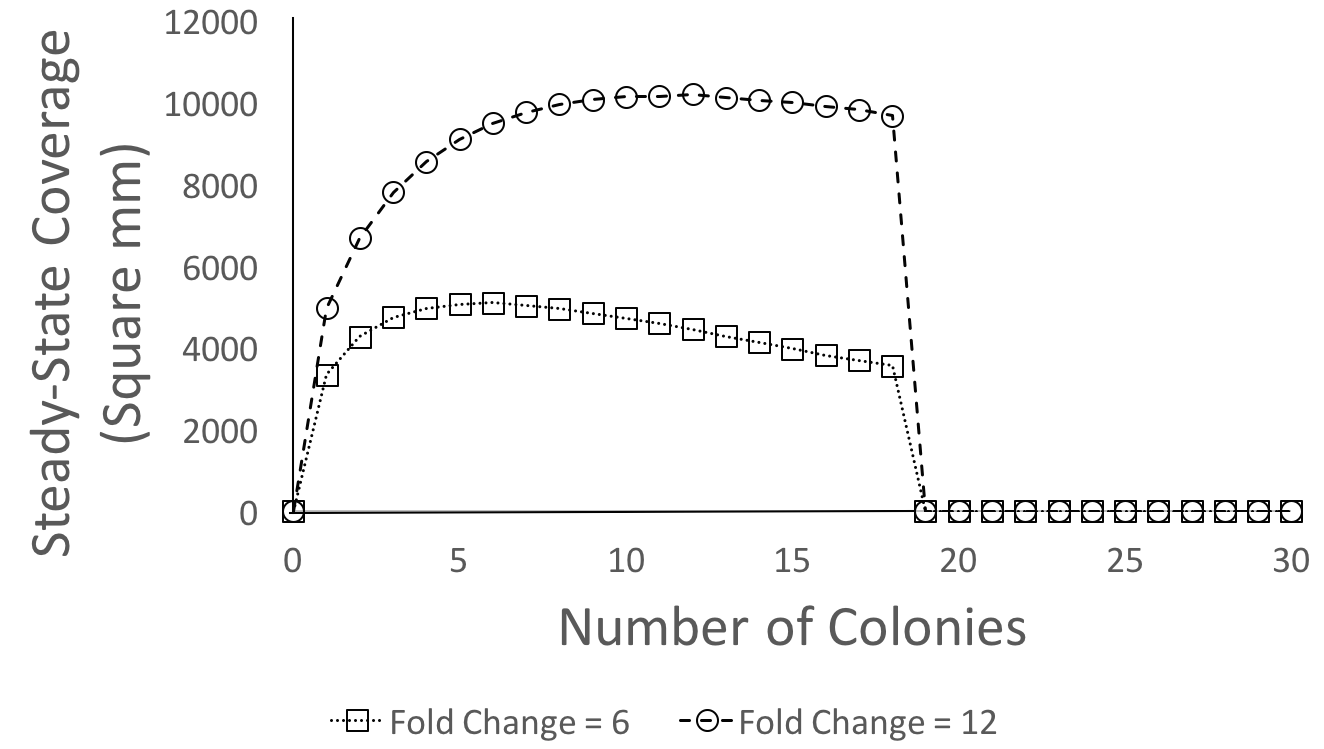}
\label{subfig:FoldChangeRobustness}}
\caption{Changes in the asymptotic coverage level curve as a function of $N_{colonies}$, in the limit of deconfined, or infinitely separated colonies for variations in (a)  sharpness of the activation (relaxing the instantaneous Heaviside-step function with a Hill curve with coefficient $b$ = 20, (b) basal production rate $r_{b}$ and (c) fold change $f$.}
\label{Fig:Robustness}
\end{figure}

To assess the robustness of our qualitative results to changes in the remaining parameters and the Heaviside approximation, we collect data for steady-state coverage in the limit of unconfined, infinitely separated colonies (Figure 3(b) in the main text). In Figure \ref{Fig:Robustness}, we compare the results generated with the default parameters used in the main text with changes in basal rate, fold change, and sharpness of the activation function. In all cases, while parameter variations shift transition points or sharpness of separation between activated and inactivated coverage levels, the existence of these transition points and separations remains intact, rendering the prediction of an LQS/GQS boundary insensitive to model parameter variations.

\subsection{Analytical Mean-Field Theory}
In this section, we will derive an analytical mean field theory to predict the LQS-GQS phase diagram, and compare these analytical results with the numerical simulations. Recall that, for long times, our time-independent, steady state solutions will satisfy
\begin{equation}
(\gamma -D \nabla^{2}) c_{AI} - r(c_{AI}) c_{cell}(\vec{x}) = 0 \ ,
\label{eq:RxnDiffEquation}
\end{equation}
where $D$ and $\gamma$ are respectively the AI signal diffusivity and decay, and $r$ is the local, AI-concentration-dependent signaling rate of the cells, given by
\begin{equation}
r(c_{AI}) = r_{b} + (r_{a}-r_{b}) \theta(c_{AI}(\vec{x}) - c_{AI_{crit}}) \ .
\label{eq:RateActivationEquation}
\end{equation}
For the system to be completely activated, we require that the concentration of $c_{AI}$ at any point with a nonzero $c_{cell}$ density be above the threshold. Note that we are keeping the total number of cells $n^{tot}_{cells}$ fixed, but dividing it out into ever more colonies $N_{colonies}$, such that there are $n^{tot}_{cells}/N_{colonies}$ cells per colony. And since we are keeping cell density of colony, $\bar{c}_{colony} $ fixed, the number of colonies $N_{colonies}$ automatically sets the long-time radius of each colony
\begin{equation}
R_{colony} = (\frac{ n^{tot}_{cells} /  N_{colonies} } {\pi \bar{c}_{colony}})^{1/2} .
\end{equation}

The remaining ambiguity concerns a choice of explicit parameterization of the long-term form of the steady state cell density $n_{cell}(\vec{x})$. In this work, we characterize $n_{cell}$ completely with the above mentioned parameters, plus one other: $L_{confinement}$, defined as the side-length of a square region of space that the cells are allowed to occupy. Then, for a given boundary size $L_{confinement}$, given colony radius $R_{colony}$ and given total number of colonies $N_{colonies}$, we consider the ensemble of all possible non-overlapping colony disk configurations that lie completely within the $L_{confinement}$ square region. 

Thus, a given system realization is characterized by the following eight independent parameters:  $\gamma$, $D$, $r_{b}$, $c_{AI_{crit}}$, $\bar{c}_{colony}$, $R_{colony}$, $N_{colonies}$ and $L_{confinemenet}$.
The two related questions that I now pose are these: for given values of the first five parameters $\gamma$, $D$, $r_{b}$,  $c_{AI_{crit}}$, $\bar{c}_{colony}$,
\begin{enumerate}
\item What is the critical colony radius $R^{crit}_{LQS}$ beyond which an individual colony, with $R_{colony} > R^{crit}_{LQS}$, will \textit{self-activate via LQS} in the limit of no neighboring influence?
\item For subcritical colony radii $R_{colony} < R^{crit}_{LQS}$, how do the parameters $N_{colonies}$ and $L_{confinement}$ set limits on the additional range of radii that allow the entire system can \textit{group-activate via GQS}?
\end{enumerate}

\subsubsection{LQS Threshold}
We begin with setting a bound on the LQS regime by calculating the steady-state concentration at the center of an isolated colony with radius $R_{colony}$ and cell density $\bar{c}_{colony}$. To start, let us just imagine that all cells are only producing at the basal rate, so that the governing equation becomes a simpler, linear equation
\begin{equation}
(\gamma -D \nabla^{2}) c_{AI} = r_{b} c_{cell}(\vec{x}) \ ,
\label{eq:RxnDiffEquationLINEARREGIME}
\end{equation}
We assume that a single-cell point source at the origin can be modeled as Dirac-Delta function, 
\begin{equation}
c^{point}_{cell}(\vec{x}) = \delta(\vec{x}) .
\label{eq:Point}
\end{equation}
Then, the exact solution in two dimensions is
\begin{equation}
c^{point}_{AI}(\vec{x}) = \frac{r_{b}}{2 \pi D} K_{0}(\frac{|\vec{x}|}{\sqrt{D/\gamma}})
\label{eq:ConcentrationPointSource}
\end{equation}
where $K_{0}$ is the zeroth-order Bessel function of the second kind. Thus, by the superposition principle, for an arbitrary cell density distribution $n_{cell}$,
\begin{equation}
c_{AI}(\vec{x}) =  \frac{r_{b}}{2 \pi D} \int d^{2} \vec{x}^{'} c_{cell}(\vec{x}')K_{0}(\frac{|\vec{x}-\vec{x}^{'}|}{\sqrt{D/\gamma}}) .
\label{eq:superposition}
\end{equation}
Thus, for an isolated colony of uniform cell density $\bar{c}_{colony}$, centered at the origin, with a radius $R_{colony}$, the AI concentration at the center in the absence of any activation would be
\begin{equation}
c^{isolated}_{AI}(0) =  (\frac{r_{b} \bar{c}_{colony}}{2 \pi D})(2 \pi) \int^{R_{colony}}_{0} dr \ r K_{0}(\frac{r}{\sqrt{D/\gamma}})
\end{equation}
which evaluates to
\begin{equation}
c^{isolated}_{AI}(0) =  \frac{r_{b} \bar{c}_{colony}}{\sqrt{D \gamma}} (\sqrt{\frac{D}{\gamma}} - R_{colony} K_{1}(\frac{R_{colony}}{\sqrt{D/\gamma}})) .
\label{eq:Isolated}
\end{equation}
Therefore, to identify the critical self-activation (via LQS) radius $R^{crit}_{LQS}$, we can set the above equation equal to the critical AI concentration $c_{AI_{crit}}$ and solve for the radius:
\begin{equation}
c_{AI_{crit}} = \frac{r_{b} \bar{c}_{colony}}{\sqrt{D \gamma}} (\sqrt{\frac{D}{\gamma}} - R^{crit}_{LQS} K_{1}(\frac{R^{crit}_{LQS}}{\sqrt{D/\gamma}})) .
\label{eq:thresholdcondition}
\end{equation}
More instructively, we can write this relation in the form 
\begin{equation}
1 - (\frac{c_{AI_{crit}} \gamma}{r_{b} \bar{c}_{colony}}) = \frac{R^{crit}_{LQS}}{\sqrt{D/\gamma}} K_{1}( \frac{R^{crit}_{LQS}}{\sqrt{D/\gamma}} ) .
\end{equation}
The parameters we used in this work were $c_{AI_{crit}} = 100 \ molecules / \mu m^{3}$, $\gamma = 0.01 / hr$, $r_{b} = 500 \ molecules / cell / hr$,  $\bar{c}_{colony} = 1 \ cell /  \mu m^{3}$, $D = 160 \ \mu m^{2} / s$ and $n^{tot}_{cells} =  \ 7000 \times 10^{3} \ cells$. Plugging in, we find $R^{crit}_{LQS} \approx 230 \ \mu m$, which yields $n^{crit}_{LQS} \approx 40$.
\subsubsection{GQS Boundary}
Now, we turn to the question of limits on GQS activation in the case where $R_{colony} < R^{crit}_{LQS}$. To answer this, we must add to the AI concentration at the center of an individual colony the contribution arising from the $N_{colonies} - 1$ neighboring colonies. This contribution is distributed over a range of all possible values that arise from the ensemble of non-overlapping disk configurations occupying the square region of size $L_{confinement}$. 
\begin{figure}
\includegraphics[scale=0.5]{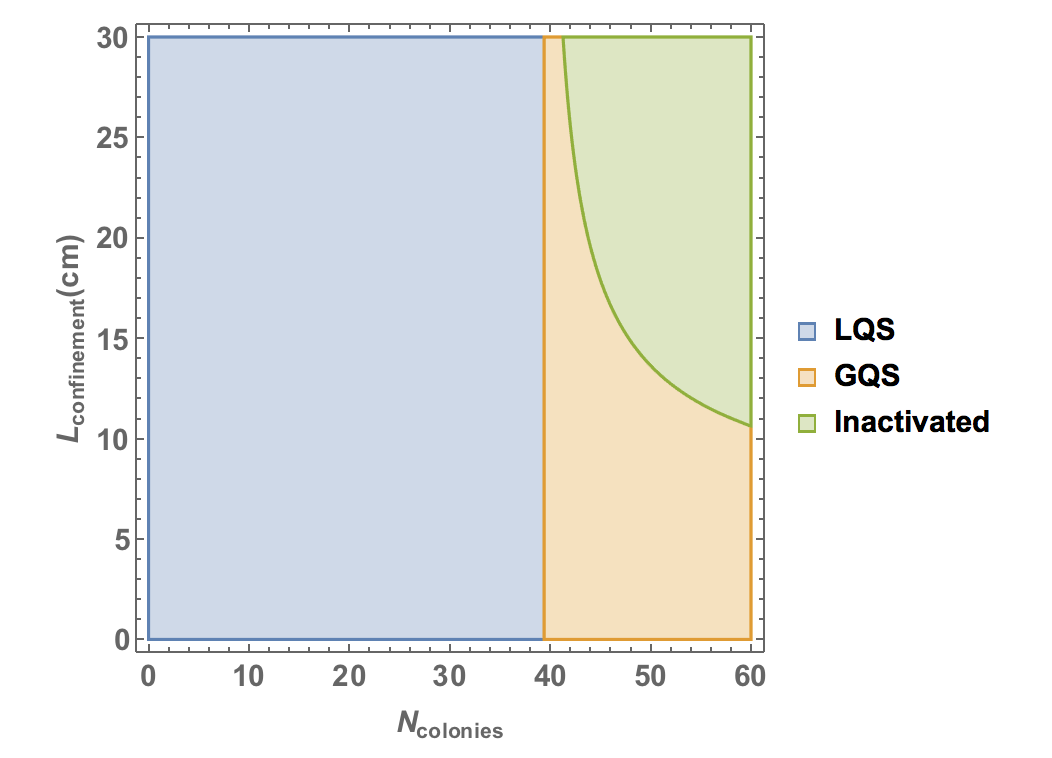}
\caption{Mean Field Phase Diagram calculated as described in this SI. The analytical results qualitatively reproduce the numerical results on the LQS/GQS transition, while the quantitative discrepancy may be attributed to the unrealistic treatment of the `near-field' source region, namely, approximation it as a Dirac-Delta point source.}
\label{Fig:MeanFieldPhaseDiagram}
\end{figure}
For simplicity, let us in this mean-field theory assume that we can coarse grain any neighbors into a continuous \textit{metapopulation} cell density, equal to the average number of cells in the confined square area, 
\begin{equation}
\bar{c}_{meta} = \frac{n^{tot}_{cells}}{L_{confinement}^{2}} .
\end{equation}
Let us also for simplicity assume that $L_{confinement} >> R_{colony}$, and that the Bessel source function decays rapidly enough such that we can effectively take $L_{confinement} \approx \infty$ for the purposes of calculation.

Then, we must modify equation (\ref{eq:Isolated}) to include not just the single-colony contribution at the center $c_{AI}(0)$, but also the neighbor contribution
\begin{equation}
c^{neighbors}_{AI}(0) \approx  (\frac{r_{b} \bar{c}_{meta}}{2 \pi D})(2 \pi) \int^{\infty}_{R_{colony}} dr \ r K_{0}(\frac{r}{\sqrt{D/\gamma}})
\end{equation}
which evaluates to
\begin{equation}
c^{neighbors}_{AI}(0) \approx  \frac{r_{b} \bar{c}_{meta}}{\sqrt{D \gamma}} (R_{colony} K_{1}(\frac{R_{colony}}{\sqrt{D/\gamma}})) .
\label{eq:Isolated}
\end{equation}

So now, if we modify our threshold condition, equation (\ref{eq:thresholdcondition}), such that the sum $c^{isolated}_{AI} + c^{neighbors}_{AI}$ equals $c_{AI_{crit}}$, we get a set of solutions ($R_{crit}$, $\bar{c}_{meta_{crit}}$) that satisfy
\begin{align}
c_{AI_{crit}} &= \frac{r_{b} \bar{c}_{colony}}{\sqrt{D \gamma}} (\sqrt{\frac{D}{\gamma}} - R^{crit} K_{1}(\frac{R^{crit}}{\sqrt{D/\gamma}})) \\
\notag
&  +  \frac{r_{b} \bar{c}_{meta_{crit}}}{\sqrt{D \gamma}} (R_{crit} K_{1}(\frac{R_{crit}}{\sqrt{D/\gamma}})).
\end{align}
Solving for this critical set ($R_{crit}$, $\bar{c}_{meta_{crit}}$), and converting the results to the equivalent in ($N_{crit}$, $L_{confinement_{crit}}$) parameter space, we get the `phase diagram' shown in Figure \ref{Fig:MeanFieldPhaseDiagram}.

\subsubsection{Discussion}
\begin{figure}
\includegraphics[scale=0.1]{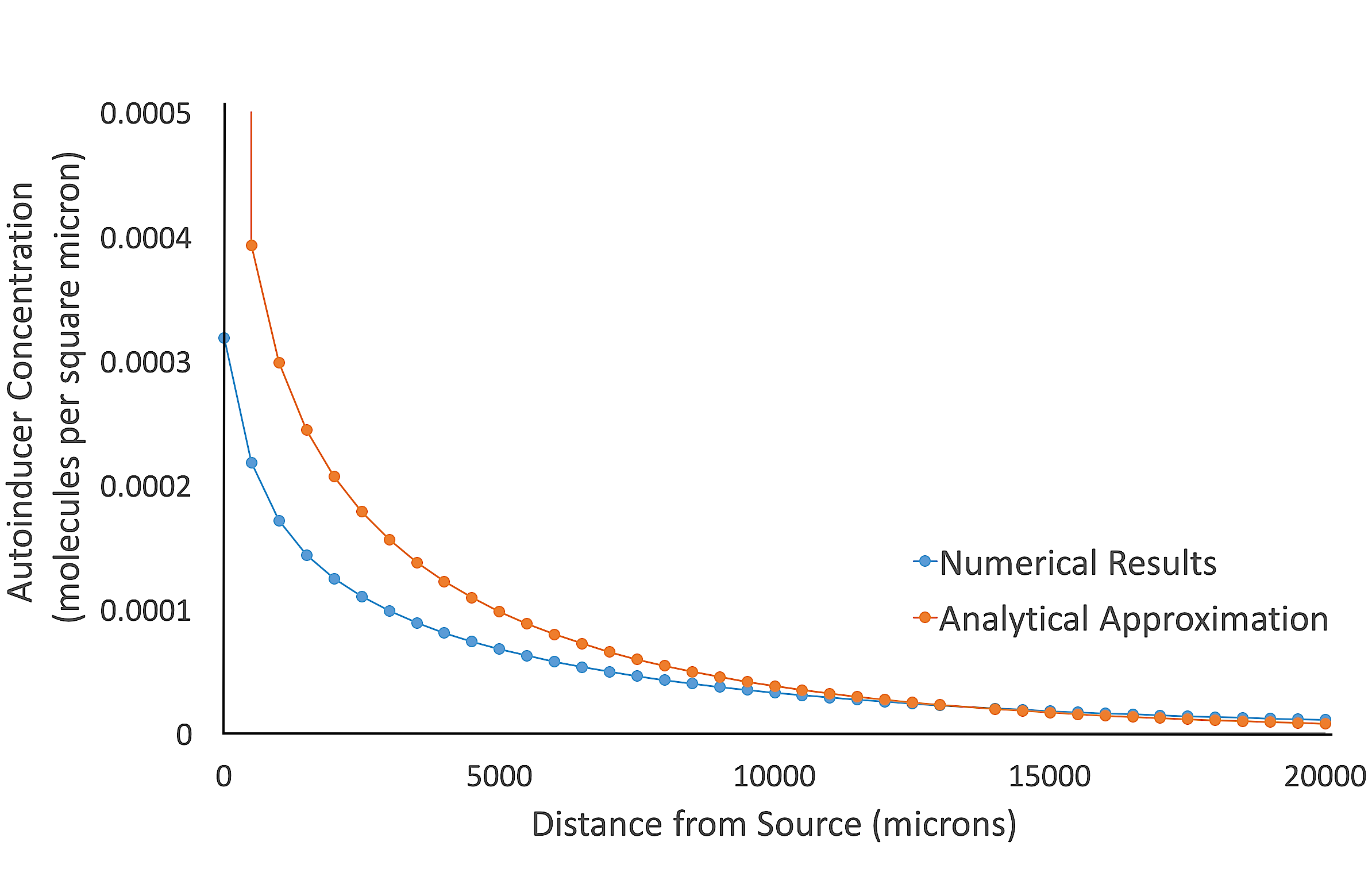}
\caption{Here, we compare the actual numerical results for the steady-state concentration due to a single cell with the the results predicted from our analytical `point-source' Green's function approximation. The unrealistic approximation of the physics close to the cell (the `near-field' region) explains the quantitative discrepancy between the two curves, and by extension, the corresponding quantitative disagreement between our analytically predicted and numerically observed phase diagrams.}
\label{Fig:SingleCellSourceFunction}
\end{figure}

As seen in the phase diagram, the essential qualitative results of the LQS/GQS transition are captured by the analytical mean-field model. However, the quantitative agreement between the analytical theory and numerical simulations is not as strong - in particular, the mean-field theory appears to overestimate the LQS boundary by a factor of two, with the resulting GQS boundaries being similarly distorted. The origin of this discrepancy can be attributed to the simplifying approximation used in equation (\ref{eq:ConcentrationPointSource}), namely, treating a finite-radius cell as a localized Dirac Delta function point source. In Figure \ref{Fig:SingleCellSourceFunction}, we compare the actual numerical results for the steady-state concentration due to a single cell with the analytical prediction of the point-source Green's function. As is seen, the two curves agree well at very long distances from the source, or in the `far-field' region. However, the analytical approximation increasingly overestimates the concentration closer to the source, or in the `near-field' region. Thus, a tentative physical interpretation of our results would be that the qualitative system behavior is well described by a far-field analytical mean field theory, with `effective' phase boundaries arising from the effects of near-field corrections. A natural future extension of this research would be in developing more accurate analytical models that quantitatively expain how near-field effects renormalize the phase diagram boundaries.

\begin{thebibliography}{99}

\bibitem{QSReview}
M. B. Miller and B. Bassler, Annual Reviews of Microbiology {\bf 55}, 165 (2001).

\bibitem{WilliamsReview} 
P. Williams, K. Winzer, W.C. Chan and M. C‡mara, Phil. Trans. of Royal Soc. B {\bf 362}, 1119 (2007).

\bibitem{JayaramanReview}
A. Jayaraman T.K. Wood, Annual Review of Biomedical Engineering {\bf 10}, 145 (2008).

\bibitem{WendellLimScience}
H. Youk and W.A. Lim, Science {\bf 343(6171)}, 1242782 (2014).

\bibitem{Hibbing}
M.E. Hibbing, C. Fuqua, M.R. Parsek and S.B. Peterson, Nature Reviews Microbiology {\bf 8}, 15 (2010).

\bibitem{Gore2016}
C. Ratzke and J. Gore, Nature Microbiology {\bf 1}, 16022 (2016).

\bibitem{Redfield2002}
R.J. Redfield, Trends in microbiology {\bf 10}, 365 (2002).

\bibitem{EffSens}
B. A. Hense, C. Kuttler, J. Muller, M. Rothballer, A. Hartmann and J.-U. Kreft, Nature Reviews Microbiology {\bf 5}, 230 (2007).

\bibitem{PerezCrosstalk2011}
P. D. PŽrez, J.T. Weiss and S.J. Hagen, BMC Systems Biology {\bf 5}, 1 (2011).

\bibitem{BoedickerAngewChem}
J.Q. Boedicker, M.E. Vincent and R.F. Ismagilov, Angew. Chem. Intl. Ed. Engl. {\bf 48(32)}, 5908 (2009).

\bibitem{QSvsDS}
S. A. West, K. Winzer, A. Gardner and S. P. Diggle, Trends in Microbiology {\bf 20}, 586 (2012). 

\bibitem{QSExpt}
M. Wu, J.W. Roberts, S. Kim, D.L. Koch, M.P. DeLisa, App. and Environ. Microbiol. {\bf72(7))}, 4987 (2006).

\bibitem{LifeIsPhysics}
N. Goldenfeld and C. Woese, Annual Review of Condensed Matter Physics {\bf 2}, 375 (2011).

\bibitem{NonEqPhysicsAndEvolution}
E. Kussell and M. Vucelja, Rep. Prog. Phys. {\bf 77}, 102602 (2014).

\bibitem{BialekCriticality}
T. Mora and W. Bialek, J. Stat. Phys. {\bf 144}, 268 (2011).

\bibitem{JeffGorePaper}
H. Celiker and J. Gore, Nature Communications {\bf 5}, 4643 (2014).

\bibitem{Boedicker2008}
C.J. Kastrup, J.Q. Boedicker, A.P. Pomerantsev, M. Moayeri, Y. Bian, R.R. Pompano, T.R. Kline, P. Sylvestre, F. Shen, S.H. Leppia, W.J. Tang and R.F. Ismagilov, Nature Chemical Biology {\bf 4(12)}, 742 (2008).

\bibitem{EvolutionArrestsInvasion}
K. S. Korolev, Phys. Rev. Lett. {\bf 115}, 208104 (2015).

\bibitem{SheepGrazing}
F. Ginelli, F. Peruani, M.-H. Pillot, H. Chate, G. Theraulaz and R. Bon, Proc. Natl. Acad. Sci. {\bf 112}, 12729 (2015).

\bibitem{LingChongYou}
S. Payne and L. You, Adv Biochem. Eng. Biotechnol. {\bf 146}, 97 (2014).

\bibitem{MehtaWingreen}
P. Mehta, S. Goyal, B.L. Bassler and N.S. Wingreen, Molecular Systems Biology {\bf 5}, 325 (2009).

\bibitem{LongWingreen}
T. Long, K.C. Tu, Y. Wang, P. Mehta, N.P. Ong, B.L. Bassler and N.S. Wingreen, PLOS Biology {\bf 7(3):e68}, 0640 (2009). 

\bibitem{DilanjiHagen2012}
G. Dilanji, J. Langebrake, P. De Leenheer and S.J. Hagen, J. Am. Chem. Soc. {\bf 134}, 5618 (2012).

\bibitem{You2015}
S. Huang, J.K. Srimani, A.J. Lee, Y. Zhang, A.J. Lopatkin, K.W. Leong and L. You, Biomaterials {\bf 61}, 239 (2015).

\bibitem{TamsirBSIMLogic}
A. Tamsir, J.J. Tabor and C.A. Voigt, Nature {\bf 469}, 212 (2010).

\bibitem{ChandlerIMSEJ}
J.R. Chandler, S. Heilmann, J.E. Mittler and P. Greenberg, The ISME Journal {\bf 6}, 2219 (2012).

\bibitem{EmereniniPLOS}
B.O. Emerenini, B.A. Hense, C. Kuttler, H.J. Eberl, PLOS ONE {\bf10(7)}, e0132385 (2015).

\bibitem{HagenTravelingWaves}
J.B. Langebrake, G.E. Dilanji, S.J. Hagen and P. De Leenheer, J. Theor. Biol. {\bf 363}, 53 (2014).

\bibitem{Powell1956}
E.O. Powell, J. Gen. Microbiol. {\bf 15}, 492 (1956).

\bibitem{StepActivation}
J. Ferkinghoff-Borg and T. Sams, Molecular BioSystems {\bf 10}, 103 (2014).

\bibitem{BSIM}
T. E. Gorochowski, A. Matyjaszkiewicz, T. Todd, N. Oak, K. Kowalska, S. Reid, K. T. Tsaneva-Atanasova, N. J. Savery, C. S. Grierson, M. di Bernardo, PLOS ONE {\bf 7}, e42790 (2012).

\bibitem{LatencyVsThroughput}
E. Grochowski, R. Ronen, J. Shen and H. Wang, Proceedings of the International Conference on Computer Design {\bf 2004}, 236 (October 2004). 

\end{thebibliography}
\end{document}